\documentclass[12pt,fleqn,reprint,aps,prl,superscriptaddress]{revtex4-1}
\usepackage{epsfig}
\usepackage{slashed}
\usepackage[intlimits]{amsmath}
\usepackage{amssymb}
\usepackage[]{hyperref}
\usepackage{color}

\begin{document}

\title{
\boldmath
\textbf{Next-to-Next-to-Leading-Order Charm-Quark Contribution to the
  $CP$ Violation Parameter $\epsilon_K$ and $\Delta M_K$}}
\unboldmath

\author{Joachim Brod}
\affiliation{Excellence Cluster Universe, Technische Universit\"at
  M\"unchen, Boltzmannstra\ss{}e 2, D-85748 Garching, Germany}
\author{Martin Gorbahn}
\affiliation{Excellence Cluster Universe, Technische Universit\"at
  M\"unchen, Boltzmannstra\ss{}e 2, D-85748 Garching, Germany}
\affiliation{Institute for Advanced Study, Technische Universit\"at
  M\"unchen, Lichtenbergstra\ss{}e 2a, D-85748 Garching, Germany}




\begin{abstract}
  The observables $\epsilon_K$ and $\Delta M_K$ play a prominent role
  in particle physics due to their sensitivity to new physics at short
  distances. To take advantage of this potential, a firm theoretical
  prediction of the standard-model background is essential. The
  charm-quark contribution is a major source of theoretical
  uncertainty. We address this issue by performing a
  next-to-next-to-leading-order QCD analysis of the charm-quark
  contribution $\eta_{cc}$ to the effective $|\Delta S|=2$ Hamiltonian
  in the standard model. We find a large positive shift of $36\%$,
  leading to $\eta_{cc} = 1.87(76)$. This result might cast doubt on
  the validity of the perturbative expansion; we discuss possible
  solutions. Finally, we give an updated value of the standard-model
  prediction for $|\epsilon_K| = 1.81(28) \times 10^{-3}$ and $\Delta
  M_K^{\text{SD}} = 3.1(1.2) \times 10^{-15}\,\text{GeV}$.
\end{abstract}


\maketitle


Strangeness-changing neutral-current transitions play an important
role in particle physics. The parameter $\epsilon_K$, measuring
indirect $CP$ violation in the neutral Kaon system, has received
increased attention recently due to the discrepancy between the
theoretical prediction and the experimental
measurement~\cite{Buras:2008nn, Lunghi:2008aa,Lenz:2010gu,
  Lunghi:2010gv}. In addition, together with the Kaon mass difference
$\Delta M_K$, it provides strong constraints on many models of new
physics.

Theoretical predictions for $\Delta M_K^{\text{SD}}$ and $\epsilon_K$
are calculated in the framework of effective field theories, which
allow to separate short- and long-distance contributions, and to sum
all terms which are enhanced by powers of large logarithms
$\log(m_c^2/M_W^2)$ using the renormalisation group. The relevant
$|\Delta S|=2$ effective Hamiltonian in the three-quark theory reads
\begin{multline}
    \mathcal{H}^{\Delta S=2}_{f=3} = \frac{G_F^2}{4 \pi^2} M_W^2 \big[
      \lambda_c^2 \eta_\textrm{cc} S(x_c) +
      \lambda_t^2 \eta_\textrm{tt} S(x_t) \\ +
      2 \lambda_c \lambda_t \eta_\textrm{ct} S(x_c,x_t) \big] b(\mu)
    \tilde Q_{S2} + \textrm{h.c.}
  \label{eq:Hlo}
\end{multline}
where $G_F$ is the Fermi constant, $\lambda_i = V_{id}^\ast V_{is}$
comprises the Cabibbo-Kobayashi-Maskawa matrix elements, and $\tilde
Q_{S2} = (\overline{s}_L \gamma_{\mu} d_L)^2 $ is the leading local
four-quark operator that induces the $|\Delta S|=2$ transition,
defined in terms of the left-handed $s$- and $d$-quark fields. The
parameter $b(\mu)$ is factored out such that
\begin{equation}
  \label{eq:bkpar}
  \hat B_K = \frac{3}{2} b(\mu) 
  \frac{
    \langle \bar K^0 | \tilde Q_{S2} | K^0\rangle}{
    f_K^2 M_K^2}\, , 
\end{equation}
where $f_K$ is the Kaon decay constant, is a renormalisation-group
invariant quantity comprising the hadronic matrix element. It can be
calculated on the lattice with high
precision~\cite{Aubin:2009jh,Aoki:2010pe,Durr:2011ap,Bae:2011ff,Colangelo:2010et}. 

The loop functions $S$ can be found, for instance,
in~\cite{Buchalla:1995vs}. The QCD and logarithmic corrections are
contained in the $\eta$ factors and are known at next-to-leading order
(NLO) for the dominant top-quark contribution
($\eta_{tt}=0.5765(65)$~\cite{Buras:1990fn}). The relative suppression
of the top-quark contribution by the small imaginary part of
$\lambda_t^2$, relevant for $\epsilon_K$, lets the charm-quark
contributions compete in size. We have already performed a
next-to-next-to-leading-order (NNLO) calculation of the charm-top
contribution ($\eta_{ct}=0.496(47)$~\cite{Brod:2010mj}). Here, we
focus on the charm-quark contribution, known until now at NLO, with a
substantial error
($\eta_{cc}=1.40(35)$~\cite{Battaglia:2003in,Lenz:2010gu}).
It multiplies $S(x_c) = x_c +
\mathcal{O}(x_c^2)$, where $x_c\equiv m_c^2/M_W^2$ and $m_c =
m_c(m_c)$ is the $\overline{\text{MS}}$ charm-quark mass. The
Glashow-Iliopoulos-Maiani (GIM) mechanism cancels a potential large
logarithm at leading order (LO).

The charm-quark contribution $\eta_{cc}$ determines the short-distance
part of the Kaon mass difference $\Delta M_K^{\text{SD}}$ and enters
$\epsilon_K$ with a negative sign. The large remaining scale
uncertainty at NLO hints at potentially sizeable NNLO corrections; we
confirm this expectation by an explicit calculation in this Letter.

Our calculation proceeds in three steps: determination of the initial
conditions of the Wilson coefficients at the electroweak scale,
renormalisation-group evolution to the charm-quark scale, and matching
onto the effective three-quark theory. The new result is the
three-loop matching condition at the charm-quark scale.

The effective Hamiltonian in the five- and four-flavour theory
relevant for $\eta_{cc}$ reads 
\begin{equation}\label{eq:lags2}
  \mathcal{H}_{f=5,4}^{\Delta S=1} = \frac{4 G_F}{\sqrt{2}}
  \sum_{i=+,-} C_{i} \sum_{q,q' =u,c}
    V_{q'd}^\ast V_{qs} Q_i^{qq'} \, .
\end{equation}
Here the current-current operators are given by $ Q_{\pm}^{qq'} =
\big( (\overline{s}_L^{\alpha} \gamma_{\mu} q_L^{\alpha}
)\otimes(\overline{q}_L^{'\beta} \gamma^{\mu} d_L^{\beta}) \pm
(\overline{s}_L^{\alpha} \gamma_{\mu} q_L^{\beta}
)\otimes(\overline{q}_L^{'\beta} \gamma^{\mu} d_L^{\alpha})\big)/2$,
where $\alpha$ and $\beta$ are colour indices, and we define the
evanescent operators in such a way that the anomalous dimension matrix
is diagonal through NNLO~\cite{Buras:2006gb, Brod:2010mj}. The GIM
mechanism cancels a contribution of the $|\Delta S| = 2$ Hamiltonian
above the charm-quark scale; we verified explicitly that mixing of
dimension-six into dimension-eight operators proportional to
$\lambda_c^2$ does not occur above the charm-quark scale.

We take the initial conditions for $C_\pm$, obtained by a NNLO
matching calculation at the electroweak scale, from
Ref.~\cite{Buras:2006gb}. The dimension-eight Wilson coefficient does
not receive a contribution at the electroweak
scale~\cite{Witten:1976kx}. The running of $C_\pm$ to the charm-quark
scale can be taken up to NNLO from~\cite{Buras:2006gb}.

\begin{figure}[t]
\centering
\includegraphics[width=\columnwidth]{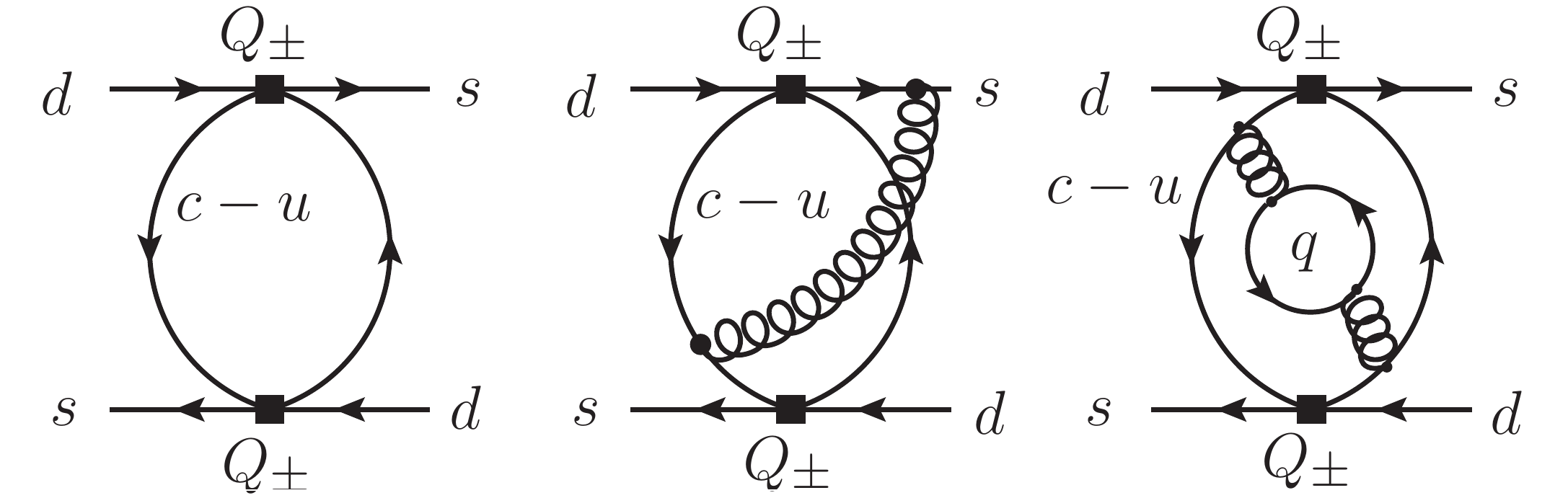}
\caption{ Sample one-, two-, and three-loop diagrams contributing to
  the matching at the charm-quark scale. Loopy lines are gluons, and
  straight lines are quarks. The combination $c-u$ arises from the GIM
  mechanism; $q$ denotes any of the quarks $u$, $d$,
  $s$. \label{fig:matchmc}}
\end{figure}

At the scale $\mu_c = \mathcal{O}(m_c)$ the charm quark is removed
from the theory as a dynamical degree of freedom. Requiring the
equality of the Green's functions in both theories at $\mu_c$ leads to
the matching condition
\begin{equation}
  \sum_{i,j=+,-} C_i C_j \langle Q_{i} Q_j
  \rangle = \frac{1}{8\pi^2} {\tilde C}_{S2}^{cc}
  \langle {\tilde Q}_{S2} \rangle \, , 
\end{equation}
which we use to determine the Wilson coefficient
$\tilde{C}_{S2}^{cc}$, defined implicitly in~\eqref{ResEta}
below. Here, angle brackets denote operator matrix elements between
$s$- and $d$-quark external states. Writing $\langle {\tilde Q}_{S2}
\rangle = r_{S2} \langle {\tilde Q}_{S2} \rangle^{(0)}$ and $\langle
Q_i Q_j \rangle = m_c^2/(8\pi^2) d_{ij} \langle {\tilde Q}_{S2}
\rangle^{(0)}$, and expanding all quantities in powers of
$\alpha_s/(4\pi)$, we find the following contributions to the
matching (a sum over $i,j = +,-$ is implied):
\begin{widetext}
\begin{equation}\begin{split}\label{eq:matchmc}
  \tilde{C}_{S2}^{cc(0)} &= m_c^2(\mu_c)
  C_{i}^{(0)} C_{j}^{(0)} d_{ij}^{(0)} \, , \\
  \tilde{C}_{S2}^{cc(1)} &= m_c^2(\mu_c) \bigg[
  C_i^{(0)} C_j^{(0)} \big(d_{ij}^{(1)} - d_{ij}^{(0)} 
  \tilde r_{S2}^{(1)}\big)
  + \big( C_i^{(1)} C_j^{(0)} +
  C_i^{(0)} C_j^{(1)} \big) d_{ij}^{(0)} \bigg] \, , \\ 
  \tilde{C}_{S2}^{cc(2)} &= m_c^2(\mu_c) \bigg[
  C_i^{(0)} C_j^{(0)} \big(d_{ij}^{(2)} - \big(d_{ij}^{(1)} - d_{ij}^{(0)} 
  \tilde r_{S2}^{(1)}\big) \tilde r_{S2}^{(1)} - d_{ij}^{(0)} 
  \tilde r_{S2}^{(2)}\big)
  + \big( C_i^{(1)} C_j^{(0)} +
  C_i^{(0)} C_j^{(1)} \big) \big(d_{ij}^{(1)} - d_{ij}^{(0)} 
  \tilde r_{S2}^{(1)}\big) \\ & + \big( C_i^{(2)} C_j^{(0)} +
  C_i^{(1)} C_j^{(1)} +  C_i^{(0)} C_j^{(2)} \big) d_{ij}^{(0)} +
  \frac{2}{3} \log\frac{\mu_c^2}{m_c^2} \bigg( \big( C_i^{(1)} C_j^{(0)} +
  C_i^{(0)} C_j^{(1)} \big) d_{ij}^{(0)} + C_i^{(0)} C_j^{(0)}
  d_{ij}^{(1)} \bigg) \bigg] \, .
\end{split}\end{equation}
\end{widetext}
The strong coupling constant $\alpha_s$ is defined in the three-quark
theory throughout this Letter, and superscripts in brackets denote the
order of the expansion in $\alpha_s$. Furthermore, we expand the
charm-quark mass defined at the scale $\mu_c$, viz. $m_c(\mu_c)$,
about $m_c(m_c)$, as in Ref.~\cite{Buras:2006gb}.

In order to evaluate the Eqs.~\eqref{eq:matchmc}, we compute the
finite parts of one-, two-, and three-loop Feynman diagrams of the
type shown in Fig.~\ref{fig:matchmc}; the evanescent operators in the
$|\Delta S|=2$ sector have been chosen as in~\cite{Brod:2010mj}. Our NLO
result confirms the calculation by Herrlich and
Nierste~\cite{Herrlich:1993yv} for the first time. The three-loop
matching calculation yields (we use the notation $\hat d_{ij}^{(2)}
\equiv d_{ij}^{(2)} - \big(d_{ij}^{(1)} - d_{ij}^{(0)} \tilde
r_{S2}^{(1)}\big) \tilde r_{S2}^{(1)} - d_{ij}^{(0)} \tilde
r_{S2}^{(2)}$; note also that $\hat d_{+-}^{(2)} = \hat
d_{-+}^{(2)}$):
\begin{multline}
 \hat d_{++}^{(2)} = \frac{1665873233}{8164800}-\frac{1573
   }{162} B_4-\frac{133}{72} D_3 +\frac{49}{36}
\zeta_2 l_c \\+\frac{4313}{216} l_c^2 -\frac{15059}{1296} l_c
+\frac{210213}{560} S_2 -\frac{1501 
  }{54} \zeta_2^2\\-\frac{7567241}{204120} \zeta_2-\frac{1697893
  }{7776} \zeta_3 +\frac{11575}{216} \zeta_4\, ,
\end{multline}
\begin{multline}
  \hat d_{+-}^{(2)}
=\frac{87537463}{1166400}+\frac{685}{162} B_4-\frac{83}{72} D_3+\frac{695}{36}
\zeta_2 l_c \\-\frac{1475}{216} l_c^2-\frac{57763}{1296}l_c-\frac{4797}{80} S_2 -\frac{791
  }{54} \zeta_2^2\\+\frac{366569}{29160} \zeta_2+\frac{57673
  }{7776} \zeta_3 -\frac{4999}{216} \zeta_4 \, ,
\end{multline}
\begin{multline}
  \hat d_{--}^{(2)}
=\frac{2129775941}{8164800} + \frac{491}{162} B_4+\frac{11}{72} D_3 
+\frac{865}{36} \zeta_2 l_c\\
   +\frac{12533}{216} l_c^2+\frac{171121}{1296}l_c+\frac{59121
    }{560} S_2  -\frac{517
  }{54} \zeta_2^2\\+\frac{9261883}{204120} \zeta_2-\frac{411709
  }{7776} \zeta_3 -\frac{7913
  }{216} \zeta_4\, ,
\end{multline}
where we defined $l_c = \log(\mu_c^2/m_c^2(\mu_c))$, $\zeta_n$ denotes
Riemann's zeta function of $n$, and the remaining constants are
defined in~\cite{Steinhauser:2000ry}. This result is new.

Since the calculation of the NNLO contributions to $\eta_{cc}$ is
quite complex, we checked our results in several ways. First of all
the calculation of the $\mathcal{O}(10\,000)$ Feynman diagrams, the
renormalisation, and the matching calculation, has been performed
independently by the two of us, using a completely different set of
computer programs, leading to identical results. On the one hand we
use qgraf \cite{Nogueira:1991ex} for generating the diagrams; the
evaluation of the integrals is then performed using the program
packages q2e/exp/MATAD~\cite{Harlander:1997zb,
  Steinhauser:2000ry}\footnote{We thank Matthias Steinhauser for
  providing us with an updated version of MATAD.}. On the other hand,
all calculations have been performed using an independent setup, based
on Feynarts \cite{Hahn:2000kx}, Mathematica, and
FIRE~\cite{Smirnov:2008iw}.

As a further check of our calculation, we verified that the matrix
elements are finite and independent of the gauge-fixing parameter
$\xi$. We have also checked analytically that $\eta_{cc}$ is
independent of the matching scales $\mu_W$, $\mu_b$, and $\mu_c$ to
the considered order of the strong coupling constant, by expanding the
full solution of the renormalisation group equations about the
respective matching scale. 

The effective Hamiltonian valid below the charm-quark threshold
contains only the single operator ${\tilde Q}_{S2}$. The
renormalisation-group evolution of the Wilson coefficient
$\tilde{C}_{S2}^{cc}$ is described by the evolution matrix
corresponding to the anomalous dimension of ${\tilde Q}_{S2}$:
\begin{equation}\label{eq:lowev}
\tilde{C}_{S2}^{cc} (\mu) = U(\mu,\mu_c) \tilde{C}_{S2}^{cc} (\mu_c) \, . 
\end{equation}
We express the coefficient $\eta_{cc}$ in a scale- and
scheme-independent way as
\begin{equation}\label{ResEta}
\eta_{cc} =
\frac{1}{m_c^2 \left(m_c\right)} \tilde C_{S2}^{cc}\left(\mu_c\right)
\left[\alpha_s^{}\left(\mu_c\right)\right]^{a_{+}}
K_{+}^{-1}(\mu_c)\, .
\end{equation}
The remaining scale dependence present in \eqref{eq:lowev} is absorbed
into
\begin{equation}
b\left(\mu\right) =
\left[\alpha_s^{}\left(\mu \right)\right]^{-a_{+}} K_{+}(\mu)\, ,
\label{DefBmu}
\end{equation}
where, up to second order in $\alpha_s$,
\begin{equation}\label{eq:J}
K_{+}(\mu) = 1 + J_{+}^{(1)}\frac{\alpha_s\left(\mu\right)}{4\pi}
  + J_{+}^{(2)} \left(\frac{\alpha_s\left(\mu\right)}{4\pi}\right)^2
\, ,
\end{equation}
and the exponent $a_{+}=2/9$ is the so-called magic number for
the operator $Q_+$ (the magic numbers as well as the matrices $J$,
comprising the higher-order QCD contributions to the
renormalisation-group evolution, are defined, for instance,
in~\cite{Gorbahn:2004my}). This scale dependence is cancelled by the
corresponding scale dependence of the hadronic matrix element, order
by order in perturbation theory. Consequently, our result is
independent of $\mu_c$ up to and including terms of
$\mathcal{O}(\alpha_s^2)$.

\begin{figure}[h]
\begin{center}
\includegraphics[width=\columnwidth]{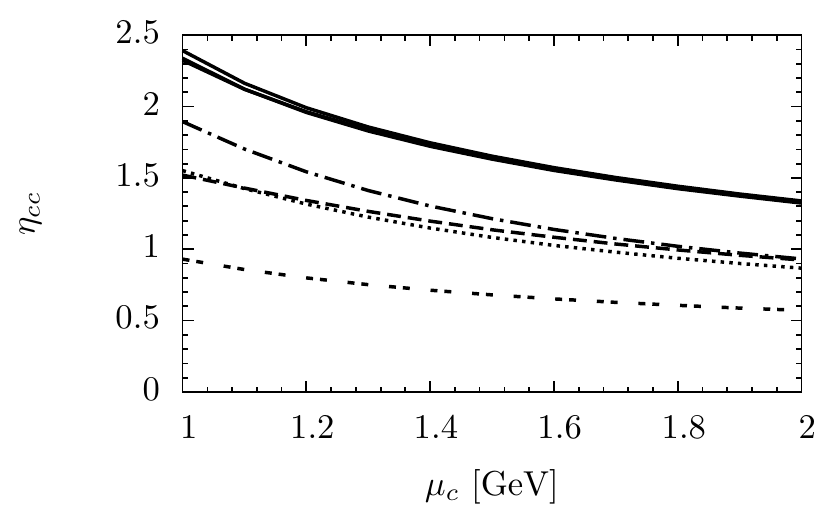}
\end{center}
\vspace{-5mm}
\caption{$\eta_{cc}$ as a function of $\mu_c$, matching at $\mu =
  m_c(m_c)$ and fixing $\mu_W=80 \, \text{GeV}$ and $\mu_b=5 \,
  \text{GeV}$. The LO result is represented by the double-dotted
  line. We also show the NLO value of $\eta_{cc}$, with the running
  $\alpha_s$ evaluated either by solving the renormalisation group
  equations numerically (dashed line), or by first computing the scale
  parameter $\Lambda_{\text{QCD}}$, either explicitly (dotted line)
  or iteratively (dash-dotted line) -- see
  Ref.~\cite{Chetyrkin:2000yt} for the details.  The resulting
  uncertainty is sizeable at NLO.  The solid lines show the
  corresponding NNLO results; now the ambiguity is almost cancelled,
  whereas the residual scale dependence is still large. }
\label{fig:mucasall}
\end{figure}

As a first estimate of the theoretical uncertainty of $\eta_{cc}$ we
study the residual scale dependence, using three different methods to
evaluate the running strong coupling
constant~\cite{Buras:2006gb}. Matching at $m_c(m_c)$ and varying
$\mu_c$ between 1 and 2 GeV (see Fig.~\ref{fig:mucasall}) and $\mu_W$
between 40 and 160 GeV we find the following numerical value at NNLO,
\begin{equation}\label{eq:eta1NNLO}
  \eta_{cc} = 1.86 \pm 0.53 _{\mu_c} \pm 0.07 _{\mu_W} \pm 0.06
  _{\alpha_s} \pm 0.01 _{m_c} \, , 
\end{equation}
where we also display the parametric uncertainties stemming from the
experimental error on $\alpha_s(M_Z) =
0.1184(7)$~\cite{Nakamura:2010zzi} and $m_c(m_c) =
1.279(13)$\,GeV~\cite{Chetyrkin:2009fv}. The dependence on the scale
$\mu_b$ and on $m_t$ is completely negligible
\footnote{Note that matching at $3\,\text{GeV}$ somewhat {\em decreases}
the central values, but {\em increases} the relative NLO and NNLO
corrections, due to the larger logarithms.}.

Varying $\mu_c$ and $\mu_W$ in the same range as above, we find at NLO
\begin{equation}\label{eq:eta1NLO}
  \eta_{cc}^{\text{NLO}} = 1.38 \pm 0.52 _{\mu_c} \pm 0.07 _{\mu_W} \pm 0.02
  _{\alpha_s} \, ,
\end{equation}
where the error indicated by the subscript ``$\mu_c$'' includes the
effect of the three ways of determining $\alpha_s$. We have included
the parametric uncertainty related to $\alpha_s$; the error
resulting from $m_c$ is negligible.

We find a substantial shift from NLO to NNLO for $\eta_{cc}$;
furthermore, we observe that the NNLO calculation does not reduce the
residual scale dependence (see Fig.~\ref{fig:mucasall}).
This reveals a bad convergence behaviour of the expansion of
$\eta_{cc}$ in $\alpha_s$, even after having summed all terms
proportional to $\alpha_s^n\log(m_c^2/M_W^2)^n$ and
$\alpha_s^{n+1}\log(m_c^2/M_W^2)^n$ using the renormalisation group.
Some of the higher-order terms leading to the residual scale
dependence are scheme dependent. In order to show that the large scale
dependence is not artificial and to gain a better understanding of its
origin, we expand the full solution of the renormalisation group
equation. We find, up to terms cubic in $\alpha_s^{}$,
\begin{flalign*}
&\eta_{cc}/(\alpha_s^{}(m_c))^{2/9}=1+\alpha_s^{}(m_c) (0.25+0.32 L_c)\\
&+\big( \alpha_s^{}(m_c) \big)^2 (1.20+0.03 L_b + 0.22 L_c + 0.27
L_c^2)\, ,
\end{flalign*}
where $L_c=\log(m_c^2/M_W^2)=-8.28$, $L_b=\log(m_b^2/M_W^2)=-5.92$,
and $\alpha_s^{}(m_c) = 0.35$ at three-loop accuracy. Here we neglect
the small terms proportional to $L_b^2$. 
This result is independent of the renormalisation scale and the
definition of the evanescent operators and depends implicitly only on
the choice of the renormalisation schemes used to determine $\hat B_K$
(cf.~Eq.~\eqref{ResEta}), $\alpha_s$, and the charm-quark mass.
The large logarithmic terms proportional to powers of $L_b$ and $L_c$
are summed to all orders by the renormalisation group in our full
result, but the constant term of the NNLO correction is more
problematic: it is almost twice as large as its NLO counterpart.
Such large constant parts are expected to lead to a large residual
scale dependence, as we indeed observe in Fig.~\ref{fig:mucasall}.

The convergence of the series can be somewhat improved by expanding
the square of the charm-quark mass multiplying $\eta_{cc}$ in
Eq.~\eqref{eq:Hlo} in powers of $\alpha_s$, noting that the
charm-quark mass receives {\em negative} corrections, although the
effect is not substantial at NNLO.

As a consequence of the discussion above, we propose the following
temporary prescription: we take $\eta_{cc}$ at $\mu_c=m_c$ as the
central value, and as the theory uncertainty the absolute size of the
NNLO correction and the residual scale dependence, added in
quadrature. This leads to
\begin{equation}
\eta_{cc} = 1.87 \pm 0.76 \, .
\end{equation}
Compared to the NLO value $\eta_{cc}^\text{NLO}$~\eqref{eq:eta1NLO},
this corresponds to a positive shift of approximately $36\%$. The
parametric uncertainty is essentially negligible with respect to the
theoretical uncertainty.

Finally we study the impact of $\eta_{cc}$ at NNLO on the prediction
of $|\epsilon_K|$ and $\Delta M_K^\text{SD}$.
We use the input values from~\cite{Nakamura:2010zzi}, in particular
$\left| V_{cb} \right|=4.06(13) \times 10^{-2}$, plus
$m_t(m_t)=163.7(1.1)$\,GeV~\cite{:1900yx},
$m_b(m_b)=4.163(16)$\,GeV~\cite{Chetyrkin:2009fv},
$\lambda=0.2255(7)$~\cite{Antonelli:2008jg},
$\kappa_{\epsilon}=0.923(6)$~\cite{Blum:2011ng},
$\xi_s=1.243(28)$~\cite{Laiho:2009eu},
$\eta_{tt}=0.5765(65)$~\cite{Buras:1990fn}, 
$\hat B_K=0.737(20)$~\cite{Laiho:2009eu}\footnote{Updates are
  available at {\tt http://latticeaverages.org/.}},
$\eta_{ct}=0.496(47)$~\cite{Brod:2010mj}, in the following formula (we
express $\bar\eta$ and $\bar\rho$ through $\sin 2\beta$; for a
discussion and definitions see~\cite{Buchalla:1995vs, Buras:2008nn}):
\begin{multline}\label{eq:eKformula}
|\epsilon_K| = \kappa_\epsilon C_\epsilon \hat B_K |V_{cb}|^2
\lambda^2 \bar \eta \big[|V_{cb}|^2(1-\bar\rho)\eta_{tt}S(x_t) \\+
\eta_{ct} S(x_c,x_t) - \eta_{cc} S(x_c)\big] \, .
\end{multline}
Using the numerical values given above, we obtain
\begin{multline}
  |\epsilon_K| = (1.81\pm 0.14_{\eta_{cc}} \pm 0.02_{\eta_{tt}}
  \pm 0.07_{\eta_{ct}} \pm 0.05_{\text{LD}} \\ \pm 0.23_{\text{parametric}})
  \times 10^{-3} \, . 
\end{multline}
The first three errors correspond to $\eta_{cc}$, $\eta_{tt}$,
$\eta_{ct}$, respectively. The error indicated by LD originates from
the long-distance contribution, namely $\hat B_K$ and
$\kappa_\epsilon$, which account for $81\%$ and $19\%$ of the
long-distance error, respectively. Half of the parametric error stems
from $|V_{cb}|$ $(49\%)$, while all other contributions are well below
$20\%$. All errors have been added in quadrature.

Compared to the prediction using the NLO value
$\eta_{cc}^{\text{NLO}}$, $|\epsilon_K^{\text{NLO}}| = 1.90(27) \times
10^{-3}$, this corresponds to a shift of approximately $-5\%$, and
overcompensates the shift of $+3\%$ found in~\cite{Brod:2010mj}. The
large perturbative corrections are thereby partially mitigated in the
observable $\epsilon_K$.

Finally we estimate the short-distance contribution to $\Delta
M_K$. Using~\cite{Buchalla:1995vs} 
\begin{equation}
  \Delta M_K^\text{SD} = \frac{G_F^2}{6\pi^2} f_K^2 B_K M_K M_W^2
  \bigg(\lambda-\frac{\lambda^3}{2}\bigg)^2 \eta_{cc} x_c
\end{equation}
we find $\Delta M_K^\text{SD} = 3.1(1.2) \times 10^{-15}
\,\text{GeV}$, where the central value accounts for $89\%$ of the
measured value. We neglected the correction due to top quarks, of the
order of 1\%. The error is dominated by $\eta_{cc}$ $(86\%)$ and $B_K$
$(6\%)$. Unfortunately, the LD contributions to $\Delta M_K$ are
poorly known; the discussion in Ref.~\cite{Bijnens:1990mz} hints at a
positive contribution. In addition, our calculation shows that also
the SD contribution cannot be computed as reliably as thought
previously, and thus the prediction of the total Kaon mass difference
suffers from large uncertainties.

We have performed the first NNLO QCD analysis of the charm-quark
contribution $\eta_{cc}$ to the $|\Delta S|=2$ effective Hamiltonian
$\mathcal{H}_{f=3}^{|\Delta S|=2}$. We confirm the analytical results
for $\eta_{cc}$ obtained at NLO in Ref.~\cite{Herrlich:1993yv} for the
first time.

The discrepancy between our standard-model prediction and the
precisely measured experimental value $|\epsilon_K|^\text{exp} =
2.228(11) \times 10^{-3}$~\cite{Nakamura:2010zzi} could be interpreted
as a tension within the standard model if we got a better control of
the theoretical uncertainty. In view of the considerable residual
scale dependence and the large NNLO shift, sizeable corrections beyond
NNLO may be expected.

Given the importance of the observable $\epsilon_K$, an effort should
be made to circumvent these difficulties. We see at least two possible
ways to proceed: in the short run, one could make use of the
cancellation of the scheme dependence between of the parameter $B_K$
and the effective Hamiltonian. One could utilize this scheme
dependence (which would affect the quantities $J$ in Eq.~\eqref{eq:J})
to achieve a better convergence of $\eta_{cc}$. Recently, new lattice
renormalisation schemes have been employed in the determination of
$B_K$~\cite{Aoki:2010pe, Aoki:2007xm}; they use nonexceptional
momentum configurations, leading to better control over lattice
uncertainties. Furthermore, they might lead to a better convergence at
NNLO, as suggested by the good perturbative behaviour of the continuum
matching for the light-quark
masses~\cite{Gorbahn:2010bf,Almeida:2010ns}. We encourage the
investigation of the effects of these schemes also on the convergence
of the series for $\eta_{cc}$, in particular, at NNLO. In the long run,
the possibility of calculating the effects of a dynamical charm quark
on the lattice might seem most promising and should be further
studied.

\begin{acknowledgments}
We thank Gerhard Buchalla, Taku Izubuchi, and Ulrich Nierste for
helpful discussions and comments on the manuscript, and Matthias
Steinhauser for providing us with numerical values of the charm-quark
mass at different orders in the strong coupling constant. JB thanks
Ulrich Nierste for suggesting to work on this topic.
\end{acknowledgments}

\bibliography{etacc}

\end{document}